USING ORDINAL VOTING TO COMPARE THE UTILITARIAN WELFARE OF A STATUS QUO AND A PROPOSED POLICY: A SIMPLE NONPARAMETRIC ANALYSIS

Charles F. Manski

Department of Economics and Institute for Policy Research, Northwestern University

December 2024

Abstract

The relationship of policy choice by majority voting and by maximization of utilitarian welfare has long been discussed. I consider choice between a status quo and a proposed policy when persons have interpersonally comparable cardinal utilities taking values in a bounded interval, voting is compulsory, and each person votes for a policy that maximizes utility. I show that knowledge of the attained status quo welfare and the voting outcome yields an informative bound on welfare with the proposed policy. The bound contains the value of status quo welfare, so the better utilitarian policy is not known. The minimax-regret decision and certain Bayes decisions choose the proposed policy if its vote share exceeds the known value of status quo welfare. This procedure differs from majority rule, which chooses the proposed policy if its vote share exceeds ½.

1. Introduction

The relationship of policy choice by majority voting and by maximization of utilitarian welfare has long intrigued a spectrum of thinkers. Riley (1990) cites commentaries in political philosophy and legal scholarship from the early 1800s onward. Pivato (2016) references studies in social choice theory from about 1970 on.

I consider an idealized scenario of choice between two policies, say A and B, in a population J. These persons have interpersonally comparable cardinal utilities [$u_j(A)$, $u_j(B)$], $j \in J$, taking values in a bounded interval, say [0, 1].[1] I assume that voting is compulsory, and that each person votes for a policy that maximizes utility.

It is evident that majority voting and maximization of utilitarian welfare need not yield the same policy choice: the former sums ordinal preferences and the latter sums interpersonally comparable cardinal utilities. Formally, let J be a probability space. To ensure uniqueness of population voting outcomes, assume that $P[u(A) = u(B)] = 0$. Then the unique policy chosen by majority voting is B if $P[u(B) > u(A)] > ½$ and is A if $P[u(A) > u(B)] > ½$. The unique policy that maximizes utilitarian welfare is B if $E[u(B)] > E[u(A)]$ and is A if $E[u(A)] > E[u(B)]$. Whether majority voting and utilitarian maximization yield the same or different policy choices is determined by the population bivariate utility distribution $P[u(A), u(B)]$.

*Example*: For each $j \in J$, let $0 < u_j(A) < 1$. Let $u_j(B)$ take the value 0 or 1. Then $P[u(B) > u(A)] = P[u(B) = 1] = E[u(B)]$. Thus, observation of the voting outcome point-identifies $E[u(B)]$ and reveals nothing about $E[u(A)]$. With no knowledge of $E[u(A)]$, the optimal utilitarian policy is unknown.  ∎

---

[1] An example in applied research is the prevalent practice in health economics of measuring health on a [0, 1] cardinal scale called a quality-adjusted life year (QALY). In a review article, Weinstein, Torrance, and McGuire (2009) write (p. S5):
> "Health states must be valued on a scale where the value of being dead must be 0, because the absence of life is considered to be worth 0 QALYs. By convention, the upper end of the scale is defined as perfect health, with a value of 1. To permit aggregation of QALY changes, the value scale should have interval scale properties such that, for example, a gain from 0.2 to 0.4 is equally valuable as a gain from 0.6 to 0.8."

Some empirical research permits persons to view sufficiently adverse health states as having value less than 0.

Notwithstanding the distinction between majority voting and utilitarian maximization, various writers have sought to connect the two procedures. The Constitutional scholar John Hart Ely wrote (Ely, 1977, p. 407):

> "democracy is a sort of applied utilitarianism . . . an institutional way of determining the happiness of the greatest number. Indeed, the connection is still more sophisticated than this. . . .By its usual reference to the greatest happiness of the greatest number, utilitarianism factors in intensities of preference: . . . But despite the bromide that democracy is blind to such intensities, it quite plainly is not. Those whose happiness is less markedly affected by a given outcome are obviously less likely to bother trying to persuade others how to vote or even for that matter to vote themselves."

Thus, Ely conjectured that when voting is not compulsory, decisions to vote will depend on intensity of preferences and tend to generate utilitarian policy choices.

In a paper titled "Condorcet Meets Bentham," the social choice theorist Marcus Pivato considered settings in which the utility distribution P[u(A), u(B)] is *concordant*, which he defined to mean that majority voting and utilitarian maximization yield the same policy choice (Pivato, 2015).[2] He recognized that utility distributions need not be concordant, but he characterized concordance as (p. 58): "a mild assumption."

Ely's conjecture and Pivato's characterization of concordance as a mild assumption optimistically suggest that society may often find it reasonable to conflate majority voting and utilitarian maximization. In contrast, the simple nonparametric analysis presented in this paper is cautionary. A fully conservative analysis, placing no restrictions on P[u(A), u(B)], yields no informative connection between the two procedures for policy choice. Constructive cautionary analysis may become feasible with assumption of arguably credible partial knowledge of P[u(A), u(B)]. I study an instructive setting in which some features of this distribution are learnable empirically rather than assumed a priori.

The setting to be considered assumes that the two policies under consideration have an asymmetric empirical status. Policy A is the status quo, and B is a proposed alternative. With this asymmetry, population outcomes under policy A are realized and may be observable, but those under B are not. I consequently assume that P[u(A)] is known but P[u(B)] is not.

---

[2] Concordance occurs if P[u(A), u(B)] is such that P[u(B) > u(A)] > ½ ⇔ E[u(B)] > E[u(A)]. Pivato expressed this in an equivalent way as sign{Med[u(B) − u(A)]} = sign{E[u(B)] − E[u(A)]}.

Whereas P[u(B)] is not observable, I suppose that a vote is taken to compare the two policies and the voting outcome observed. The voting mechanism may, for example, be an advisory referendum of the population. I make the idealized assumption that voting is compulsory and that all persons vote for the policy that maximizes utility. Then observation of the voting outcome yields knowledge of P[u(B) > u(A)].[3]

Combining the above yields empirical knowledge of P[u(A)] and P[u(B) > u(A)]. The analysis in Section 2 goes further and assumes knowledge of the joint distribution P{u(A), 1[u(B) > u(A)]}. This distribution is learnable empirically if one can jointly observe the voting behavior and status quo outcomes of the population. Important findings in Section 3 require only knowledge of E[u(A)] and P[u(B) > u(A)].

The analysis is in two parts. Section 2 uses knowledge of P{u(A)], 1[u(B) > u(A)]} to obtain a sharp bound on E[u(B)]. The analysis extends a simple derivation performed in Manski (1990) in a different context. The midpoint of the bound is ½E[u(A)] + ½P[u(B) > u(A)], whose computation only requires knowledge of E[u(A)] and P[u(B) > u(A)], not complete knowledge of P{u(A), 1[u(B) > u(A)]}. The bound on E[u(B)] is generally informative, but it necessarily contains the known value of E[u(A)]. Hence, one cannot conclude whether policy A or B yields higher utilitarian welfare.

Section 3 draws implications for utilitarian planning. I consider a planner who can bound E[u(B)] as described above. Given that the bound on E[u(B)] contains E[u(A)], the planner faces a problem of decision making under uncertainty. I address this problem as I have in Manski (2024) and elsewhere, considering Bayesian planning, the maximin criterion, and the minimax-regret (MMR) criterion.

I find that a maximin planner always chooses A. The Bayesian policy choice compares E[u(A)] with the subjective mean of E[u(B)]. If the subjective mean for E[u(B)] is the midpoint of the bound derived in

---

[3] These idealized assumptions can be relaxed to some degree without materially affecting the finding to be obtained. For example, rather than assume that voting is compulsory, it would approximately suffice for a large random sample of the population to vote.

Section 2, the unique Bayes choice is B if $P[u(B) > u(A)] > E[u(A)]$ and is A if $P[u(B) > u(A)] < E[u(A)]$. I find that this is the unique MMR choice as well.

The findings on Bayesian and MMR policy choice provide an instructive contrast with majority voting. Majority voting chooses B if $P[u(B) > u(A)] > ½$ and A if $P[u(B) > u(A)] < ½$. The welfare realized with the status quo policy is disregarded. To the contrary, the present analysis shows that the MMR (and sometimes the Bayesian) voting threshold for choice of the proposed policy over the status quo is the magnitude of status quo welfare.

2. Bounding E[u(B)]

The bound on E[u(B)] is easily obtained by application of the Law of Iterated Expectations. Proposition 1 gives the result.

*Proposition 1*: Consider choice between policies A and B in population J, a probability space. Let members of J have interpersonally comparable cardinal utilities $[u_j(A), u_j(B)]$, $j \in J$, taking values in [0, 1]. Assume that $P[u(A) = u(B)] = 0$, voting is compulsory, and each person votes for a policy that maximizes utility. Then knowledge of $P\{u(A), 1[u(B) > u(A)]\}$ yields this sharp bound on E[u(B)]:

(1) $E[u(A)|u(B) > u(A)] \cdot P[u(B) > u(A)] < E[u(B)]$

$$< P[u(B) > u(A)] + E[u(A)|u(A) > u(B)] \cdot P[u(A) > u(B)].$$

Proof: By the Law of Iterated Expectations,

(2) $E[u(B)] = E[u(B)|u(B) > u(A)] \cdot P[u(B) > u(A)] + E[u(B)|u(A) > u(B)] \cdot P[u(A) > u(B)]$.

E[u(B)|u(B) > u(A)] and E[u(B)|u(A) > u(B)] are not known. However, the event {u(B) > u(A)} holds if and only if u(A) < u(B) ≤ 1 and the event {u(A) > u(B)} holds if and only if 0 ≤ u(B) < u(A). Hence, E[u(A)|u(B) > u(A)] < E[u(B)|u(B) > u(A)] ≤ 1 and 0 ≤ E[u(B)|u(A) > u(B)] < E[u(A)|u(A) > u(B)]. Inserting these bounds in (2) yields the bound in (1).

Q.E.D.

Studying a different context, Manski (1990) derived bound (1) in the special case where $u_j(A)$ is constant across j ∈ J. Then (1) reduces to

(1′)  u(A)·P[u(B) > u(A)]  <  E[u(B)]  <  P[u(B) > u(A)] + u(A)·P[u(A) > u(B)].

In this earlier work, the objective was to learn the mean subjective probability that members of a population place on a specified uncertain event. It was assumed that subjective probabilities are not observed, but persons give yes/no responses to a question asking whether the event will occur. It was assumed that a person responds "yes" if his subjective probability is above a known threshold and "no" if it is below the threshold. In the present notation, $u_j(B)$ is person j's subjective probability and u(A) is the threshold separating the yes/no responses. It may be credible to assume that everyone uses the threshold u(A) = ½, responding "yes" if and only if the subjective probability is above ½.

Bound (1) is generally informative; that is, a proper subset of [0, 1]. An exception occurs when $u_j(A)$ takes the value 0 or 1 for almost every j ∈ J. Then E[u(A)|u(B) > u(A)] = 0 and E[u(A)|u(A) > u(B)] = 1. It follows that (1) is the trivial bound [0, 1].

Although bound (1) is generally informative, it always contains E[u(A)] in its interior. Hence, it does not reveal whether policy A or B has the larger utilitarian welfare. To see this, apply the Law of Iterated Expectations to E[u(A)], obtaining

(3)  E[u(A)] = E[u(A)|u(B) > u(A)]·P[u(B) > u(A)] + E[u(A)|u(A) > u(B)]·P[u(A) > u(B)].

All quantities on the right-hand side are in the interval [0, 1]. Hence, the lower bound in (1) is weakly less than E[u(A)] and the upper bound in (1) is weakly greater than E[u(A)].

The midpoint of bound (1) will play an important role in the analysis of Section 3. The midpoint is

(4)  ½{E[u(A)|u(B) > u(A)]·P[u(B) > u(A)] + P[u(B) > u(A)] + E[u(A)|u(A) > u(B)]·P[u(A) > u(B)]}

= ½E[u(A)] + ½P[u(B) > u(A)].

Observe that the expression on the right-hand side of (4) is a function only of E[u(A)] and P[u(B) > u(A)]. The conditional expectations E[u(A)|u(B) > u(A)] and E[u(A)|u(A) > u(B)] are not germane. Thus, complete knowledge of the joint distribution P{u(A)], 1[u(B) > u(A)]} is not needed to determine the midpoint of the bound. It suffices to know the utilitarian welfare of the status quo policy and the fraction of the population who vote for the proposed policy.

3. Policy Choice under Uncertainty

Now consider a planner who uses (1) to bound E[u(B)]. The bound contains E[u(A)], making utilitarian maximization infeasible. Nevertheless, decision theory provides various reasonable decision criteria. I address this problem of social planning under uncertainty as I have in Manski (2024) and elsewhere, considering Bayesian planning, the maximin criterion, and the minimax-regret criterion.

3.1. Bayesian Policy Choice

A Bayesian planner places a subjective distribution on E[u(B)], say π, whose support is a subset of bound (1), and computes the subjective mean E_π[u(B)]. The planner chooses B if E_π[u(B)] > E[u(A)] and A if E_π[u(B)] < E[u(A)].

The Bayesian policy choice simplifies if $E_\pi[u(B)]$ is the midpoint of bound (1), given in (4). The difference between the midpoint of the bound and $E[u(A)]$ is $\tfrac{1}{2}P[u(B) > u(A)] - \tfrac{1}{2}E[u(A)]$. Hence, a Bayesian planner whose subjective mean for $E[u(B)]$ is the midpoint of the bound chooses B if $P[u(B) > u(A)] > E[u(A)]$ and chooses A if $P[u(B) > u(A)] < E[u(A)]$.

3.2. Maximin Policy Choice

A planner who does not find it credible to place a subjective distribution on $E[u(B)\}$ may use the maximin criterion. It was shown in Section 2 that the lower bound on $E[u(B)]$ is less than the known value $E[u(A)]$. Hence, the maximin policy choice is always A, regardless of the empirical evidence in $P\{u(A), 1[u(B) > u(A)]\}$.

3.3. Minimax-Regret Policy Choice

Alternatively, a planner who does not find it credible to place a subjective distribution on $E[u(B)\}$ may use the MMR criterion. Policy A yields positive regret if B gives higher utilitarian welfare than A, and vice versa. The maximum regret for A occurs if $E[u(B)]$ is at its upper bound, whereas the maximum regret for B occurs if $E[u(B)]$ is at it lower bound. Thus, the maximum regrets for A and B respectively are

(5a) $P[u(B) > u(A)] + E[u(A)|u(A) > u(B)] \cdot P[u(A) > u(B)] - E[u(A)] =$

$$P[u(B) > u(A)] - E[u(A)|u(B) > u(A)] \cdot P[u(B) > u(A)],$$

(5b) $E[u(A)] - E[u(A)|u(B) > u(A)] \cdot P[u(B) > u(A)] = E[u(A)|u(A) > u(B)] \cdot P[u(A) > u(B)].$

The MMR choice is A if (5a) is less than (5b), and it is B if (5b) is less than (5a). Inspection of (5a) and (5b) shows that A is the MMR choice if $P[u(B) > u(A)] < E[u(A)]$, and it is B if $P[u(B) > u(A)] > E[u(A)]$. Thus, the MMR choice is the same as that of a Bayesian planner whose subjective mean for $E[u(B)]$ is the midpoint of bound (1).

4. Conclusion

Although obtained in an idealized scenario, the findings on Bayesian and MMR policy choice in Section 3 suggest a broad practical implication for evaluation of majority voting as a mechanism for choice between a status quo and a proposed policy. Majority voting chooses B if $P[u(B) > u(A)] > ½$ and A if $P[u(B) > u(A)] < ½$. The welfare realized with the status quo policy is disregarded. Status-quo welfare is also disregarded with voting procedures that require a specified supermajority, say 3/5 or 2/3, to replace a status quo with a new policy. I am not aware of existing voting procedures that make the threshold for choice of a new policy increase with the cardinal welfare attained by the status quo, as found in Section 3. Making the threshold increase with status quo welfare warrants serious consideration in societies that aim to maximize utilitarian welfare.